# On Measuring Available Capacity in High-speed Cloud Networks

Ganapathy Raman Madanagopal, Christofer Flinta, Andreas Johnsson, Farnaz Moradi and Daniel Turull
Research Area Cloud Technologies
Ericsson Research
Stockholm, Sweden
(ganapathy.raman.madanagopal, christofer.flinta, andreas.a.johnsson, farnaz.moradi, daniel.turull)@ericsson.com

*Abstract*—Measurement of available path capacity with high accuracy over high-speed links deployed in cloud and transport networks is vital for performance assessment and traffic engineering. Methods for measuring the available path capacity rely on sending and receiving time stamped probe packets. A requirement for accurate estimates of the available path capacity is the ability to generate probe packets at a desired rate and also time stamping with high precision and accuracy. This is challenging especially for measurement systems deployed using general purpose hardware. To touch upon the challenge this paper describes and evaluates four approaches for sending and receiving probe packets in high-speed networks (10+ Gbps). The evaluation shows that the baseline approach, based on the native UDP socket, is suitable for available path capacity measurements over links with capacities up to 2.5 Gbps. For higher capacities we show that an implementation based on Data Plane Development Kit (DPDK) gives good results up to 10 Gbps.

I. INTRODUCTION

With the increase of Internet traffic volumes it is vital for cloud providers to know how congested the different links are in their networks. Around 70 % of all companies are using or exploring cloud services today, but 45 % consider capacity requirements as a barrier to cloud adoption [1]. In order to be able to meet these requirements the cloud providers need to monitor the available capacity of links in the datacenter, between datacenters and to the users.

Typical link capacities in these networks are currently at 10 Gbps and the trend is to increase these capacities. Monitoring link usage at 10 Gbps is a huge challenge since time stamping of the transmitted packets needs to be done accurately with high precision. Also, storing time stamps and packets and collecting the information by a management system is a demanding task, requiring large CPU, memory and network resources. Since monitoring systems using passive probes puts high requirements on the network nodes and require all nodes to have these probes, there is a rising interest for methods using active measurements for capacity estimation.

The basic principle for active capacity measurements is to inject probe packets into the network from a sender to a receiver with specific time intervals between the packets and analyze how these time intervals change during the transport in a network path. Examples of active probing methods are BART (Bandwidth Available in Real-Time) [2], Pathload [3] and Pathchirp [4], which all estimate available path capacity (APC), defined as the unused part of the capacity of the tight link in a network path [5]. For a more complete picture of methods in the area we refer to a survey paper by Chaudhari and Biradar [6].

Typically, an APC method sends a sequence of packet trains, where a packet train is a group of packets with a specified sending time interval between each packet. These packets are time-stamped at both the sender and the receiver. The scheme of time intervals of a train is method dependent. Some methods use different time intervals between each packet in the train, e.g. using increasing time intervals [4], while other methods use the same time interval between each packet in the train, but vary that time interval for different trains [2]. The time intervals and the number of bits per packet can be translated to a momentary rate for each packet-pair in the train or as a rate for the whole train. This rate can be calculated at the sender as the send rate and at the receiver as the receive rate. By comparing the send rate with the receive rate it can be deduced whether there is congestion in the network. The send and receive rate can also be used as input to an algorithm for estimating APC. The length of the packet trains also depends on the method, where some methods use large trains of thousands of packets [3], while other methods use short trains in the order of tens of packets [2] [4]. All methods rely on the ability to generate probe packet trains at a required rate and time-stamp the packets with high precision, which becomes a challenging task for 10 Gbps links. Recent work [7] shows the difficulties for time stamping the packets at high speeds.

In this paper we investigate four different methods for sending and receiving packet trains and evaluate their performance in a high-speed network. Each method has been implemented with a sender and a receiver and tested in a lab network. While other work has shown that it is possible to generate long sequences of packets at 10 Gbps, we investigate here the possibilities to send and receive very short trains of 10 – 100 packets per train at this speed. The scope of the work has been to look into how fast the different methods can generate these short packet trains and how accurate the methods can measure the send and receive rates for the trains. We have limited the tests to a network with an empty link of 10 Gbps.

The rest of the paper is organized as follows. Section 2 outlines the challenges targeted by this paper. Section 3

describes our testbed for experiments. Section 4 reviews technologies for accurate time stamping and high-speed measurements. Section 5 provides an evaluation and insights on using the methods, while section 6 provides a discussion. The paper ends with section 7 and conclusions.

## II. PROBLEM STATEMENT

Most methods for estimating available capacity rely on sending packet trains at a precise rate with accurate time stamping at both the sender and receiver. This can be summarized as three main requirements for all methods that depend on self-induced congestion in the network: (1) the sender must be able to send probe packets with a rate corresponding to the link capacity of the tight link in a network path, (2) the sender must be able to estimate the actual send rate for each packet train by reading the time stamps of the sent packets accurately, and (3) the receiver must be able to estimate the actual receive rate for each packet train by reading the time stamps of the received packets accurately.

These requirements can be hard to fulfil in a high-speed network with link capacities of 10 Gbps or higher. The first requirement deals with the task to generate packets with rates at least 833 kpps using 1500-byte packets. This means that the packets need to be generated with a maximum time between packets of around 1 microsecond. The sender and receiver applications running in user space are often built to communicate with the host network stack using the send() or recv() system calls on UDP sockets, which works well for networks with limited link speeds. For higher rates the processing time in the IP stack may take longer than the required send time between packets, which means that it may not be possible to achieve the desired send rate.

The second requirement deals with the task to read the timestamps of the sent measurement packets. If these time stamps are read in user space they become obsolete when the packets have reached the NIC after passing through various layers in the network stack with multiple queues. The time between reading a time stamp and the actual sending of a packet can also be randomly delayed by interrupts or process scheduling. In a high-speed network these distorted time stamps then affect the estimated send rate so it will not reflect the actual send rate, which makes it hard to use for capacity estimations.

The third requirement deals with the task to read the timestamps of the received measurement packets. Time stamping is affected at the receiver in the same way as in the sender concerning interrupts and process scheduling. In addition to this, packets are often buffered in the receiver before they are time stamped. This may result in an artifact where the estimated receive rate appears to be higher than the actual receive rate, which in turn gives erroneous calculation of the APC.

The challenge for capacity measurements in a high-speed network is thus to make sure that the differences between the desired send rate, the estimated send rate and the actual send rate are sufficiently small on the sender side. Similarly, on the receiver side the difference between the estimated receive rate and the actual receive rate must be sufficiently small.

First, confirm that you have the correct template for your paper size. This template has been tailored for output on the US-letter paper size. If you are using A4-sized paper, please close this file and download the file "MSW_A4_format".

## III. TESTBED

To investigate the challenges we have designed a testbed with two nodes connected via a single-hop 10 Gbps link. The nodes are identical with the following hardware configuration. Each node is equipped with 24 Intel Xeon CPUs with base frequency of 2.8 GHz. The operating system is Ubuntu 14.04 with Linux Kernel version 3.18.4. Note that the experiments only utilized one CPU on each node.

The two nodes execute an in-house designed sender and receiver, using different methods for sending and receiving. All packets from the sender are grouped into trains with specific lengths, N, and transmitted with different rates up to 10 Gbps. For each train only the first and the last packet is time stamped just before sending. The send rate for each train can then be calculated by dividing the number of bits at the Ethernet layer for the first N-1 packets by the time difference between the first and the last packet. On the receiver side each packet is time stamped just after being received. The receiver stores all packets for each train before calculating the receive rate, which is calculated in a similar way as the send rate, using the last N-1 packets and the time difference between the first and last received packet.

The packets are formatted with TWAMP [8] fields, even if some methods do not reflect packets or use these fields, but only record time stamps at the sender and receiver. The methods that do reflect packets are implemented with the extension described in RFC 6802 [9], where each packet train is buffered at the reflector before sent back to the receiver. In this way we minimize the impact on time stamping by the reflecting process. Note that these reflected trains are not used for capacity estimation for the reverse path in these experiments; we consider the problem to be the same in both directions and therefore focus on only the forward path.

## IV. METHODS FOR HIGH-SPEED MEASUREMENTS

We have selected four methods to be implemented and tested based on different architectures, including native UDP as a baseline method; we name them nativeUDP, LibPcap, Netmap and DPDK. For each method a TWAMP sender and reflector is implemented and executed on the testbed. Due to page limitations we leave out the details and mainly focus on a brief overview of the methods. Note that the methods primarily focus on the send and receive functionality of a TWAMP application. The main ideas are to cut packet process time by creating packet headers in advance, bypassing the IP stack and using polling and spin-loops for time stamp precision.

The nativeUDP method uses the system calls sendto() and recvfrom() to send and receive packets over the IP stack. All time stamps are obtained using traditional system calls in user space with nanosecond precision. This method is used by most measurement methods and is the base-line method used for comparison in this paper.

The LibPcap method utilizes the pcap API [10] for sending and receiving network traffic over a raw socket at the user level in Linux. Packets are sent using sendpacket() while pcap_loop() is repeated over the receiving socket to receive and deliver the packet. Time stamping is performed in user space.

The third method, Netmap, is based on a novel framework from Luigi Rizzo et.al [11] which aims to reduce the cost of moving traffic between hardware and the host network stack by partially disconnecting the NIC from the host stack. A shared memory is defined to exchange packets between the NIC and the application, and the application uses the poll() mechanism to send and receive packets.

The Data Plane Development Kit (DPDK) [12] is used for the fourth method. It defines a set of libraries and NIC drivers for faster processing of packets in data plane applications, without the use of the native network stack. It provides a framework for building high-speed packet networking applications over Intel x86 processors, through the Environment Abstraction Layer (EAL). Unlike Netmap, which takes control of the interface only when a user application is using the Netmap API, DPDK takes full control of the NIC once the NIC is bound to it. Further, it requires that at least one core of a CPU is dedicated for DPDK packet processing.

In all four implementations the following parameters are transferred at session start from the main application to the TWAMP sender: reflector IP and port, number of trains, number of packets per train and desired send rate for the trains. In the DPDK implementation two extra parameters are supplied: number of dedicated CPU cores and number of memory channels to be associated with the TWAMP application.

Further, memory for N packets is pre-allocated in order to avoid allocating memory dynamically on per-packet basis. In the nativeUDP, LibPcap and Netmap implementations the memory is allocated in user space, while in the DPDK implementation the memory is allocated in kernel space.

In all implementations the TWAMP fields are filled in advance for all packets in a train except for the sending timestamp, which is filled in just before sending each packet. UDP, IP and MAC headers are also calculated and filled in advance by the LibPcap, Netmap and DPDK implementations, while the nativeUDP implementation rely on the host IP stack for packet headers. Further, a spin-loop with nano-second precision is used for creating delays for all implementations.

The reported metrics from the implementations are the estimated send rate and the estimated receive rate, while the desired send rate is an input parameter. The actual send and receive rates are not measured directly, but rather inferred by using the DPDK method, see details below.

## V. EVALUATION

In order to evaluate the implemented methods, we conduct four different sets of tests in the testbed. In the first set of experiments, the four methods are tested one by one, with the same approach in both sender and receiver. Packet trains are transmitted as fast as possible from the sender, and received at the other end. The train size is set to 50 packets and the packet size is 1514 bytes, including Ethernet, IP, UDP, and TWAMP headers. In order to obtain reliable results, the same experiment is repeated 10 times for each method.

The second set of experiments only applies for the DPDK method, where the tuning parameters are the train size and the desired send rate. The packet size is again 1514 bytes, and the experiments are repeated 10 times. The third and fourth sets of experiments use DPDK as either sender or receiver and each of the other methods on the opposite node. All four sets are described in detail below. Note that the theoretical maximum rate at the Ethernet layer with packets of 1514 bytes over a 10 Gbps link is only 9.87 Gbps, since the preamble parts and the inter-frame gaps are not counted.

### A. Estimated send and receive rates for each method

The aim of the first set of experiments is to send packet trains with the desired send rate of 10 Gbps and see if any of the methods can send and receive trains at this rate. The left part in Fig. 1 shows the estimated send rates for the four methods, as reported from the application. The nativeUDP method reports that it can generate packet trains at around 2.5 Gbps, LibPcap at 6 Gbps, while both Netmap and DPDK claim to send at near wire speed with very low variance. As a conclusion from the results for the Sender side it is clear that Netmap and DPDK are the only two candidates for sending packets at 10 Gbps.

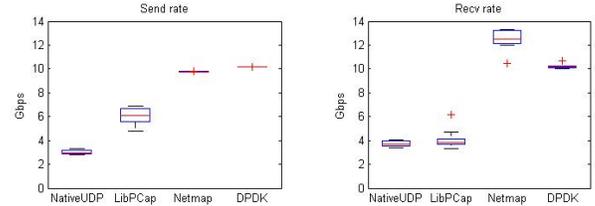

Fig. 1. Estimated send rates (left) and estimated receive rates (right) for the four methods

The right part of Fig. 1 shows the estimated receive rates for the four methods, as reported from the application. Only DPDK seems to report send rates at wire speed with low variance. Netmap shows receive rates over 12 Gbps with high variance, which indicates packet buffering before first time stamps. NativeUDP and LibPcap both show receive rates around 4 Gbps.

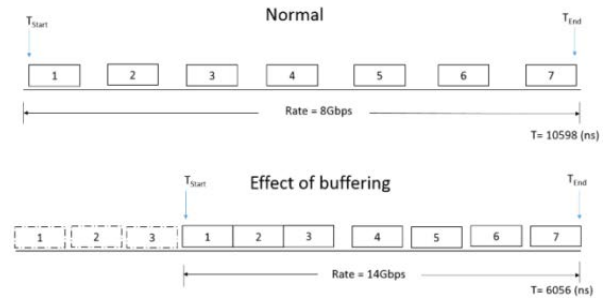

Fig. 2. Buffering of packets delays time stamping, which shortens the estimated receive time.

Fig. 2 illustrates the reason for higher rates seen at the receiving side when using Netmap as the architecture. Netmap uses a polling mechanism and the overhead for the system calls are amortized over large batches. Due to buffering, the inter-packet gap is lost and many of the packets contain the same timestamps. Thus the calculated time difference between the first and the last packet will be much smaller than the original time difference which leads to much higher estimated receiving rates.

Taking into account the results from both the sender and the receiver from this first set of experiments it seems that DPDK is the only method that is able to generate packet trains at 10 Gbps and report the correct send and receive rates. The low variance of the transmission rates at both the sender and the receiver further supports this.

Since we have no monitoring of the actual rates we cannot claim that any of the methods are showing the correct values for the estimated send and receive rates. However, it has been shown [13] that DPDK can be used to generate packets with rates over 10 Gbps, indicating that the actual send and receive rate in our experiments really reflects the desired send rate. To further validate the DPDK method we perform more detailed tests, described in the next section.

### B. Train length and rate impact on DPDK

In the second set of experiments we use DPDK as both sender and receiver. We vary the desired send rate and the train length and calculate the estimated send and receive rates. The results are summarized in Fig. 3.

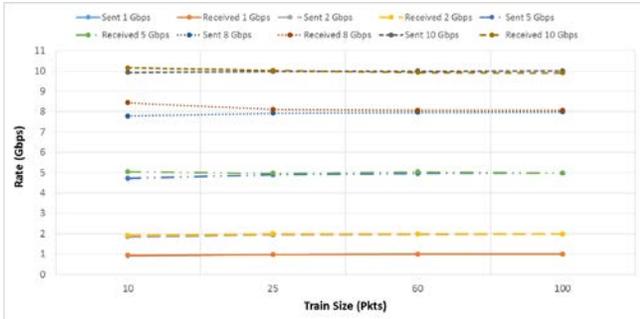

Fig. 3. Estimated send and receive rates for different train lengths and different desired send rates, using DPDK as both sender and receiver.

There are three observations to be made. First, for all experiments the estimated receive rate is larger than the corresponding estimated send rate, with the desired rate mostly in between. Second, the gap between the estimated send and receive rate tends to decrease with increasing train size, where both the send and receive rate are asymptotically going towards the desired rate. Third, this gap is very low for low desired send rates – even for short trains.

The reason for this behavior is not fully clear, but a higher receive rate than expected for shorter trains indicates that there is some buffering of the first packet in the receiver. This lag in time stamping gives then a lower calculated receive time for the train, see Fig. 4, which in turn translates to a higher estimated receive rate. For longer trains the relative error between the estimated and the actual time decreases, which corresponds to decreased error between estimated and actual receive rates, explaining the second observation. Similarly, the relative error decreases for lower rates, since the receive time for a train is longer for lower rates, explaining the third observation.

The lower estimated send rate than expected for shorter trains indicate that there is some delay between sending the last packet and getting the time stamp for this packet, which gives a higher calculated send time than the actual send time. This send time translates into a lower estimated send rate than expected. In the same way as for the receive rate this error decreases with longer trains and lower desired sending rates.

This convergence of estimated send rate, estimated receive rate and desired send rate for longer trains for all tested desired send rates strengthens the view that DPDK estimates send and receive rates with high accuracy. Our conclusion is that DPDK is a reliable method for estimating link capacities for empty links up to at least 10 Gbps.

### C. Comparing Sender methods with DPDK as Receiver

With the above conclusion of DPDK as an accurate method for sending and receiving packet trains up to at least 10 Gbps, we decide to use it as a tool for examining the other methods in more detail. The third set of experiments is using DPDK as a receiver and looks into the estimated and actual send rates for each of the four methods as the sender. The idea is that the estimated receive rate reported by DPDK can be used as an approximation of the actual receive rate, which in turn equals the actual send rate from the sender, since the link is empty. The estimated send rates from the other methods can then be compared to the actual send rates as reported from the DPDK receiver. The desired send rate is 10 Gbps for all experiments.

TABLE 1. ESTIMATED (EST) AND ACTUAL (ACT) SEND RATES IN GBPS FOR DIFFERENT SENDER METHODS, USING DPDK AS RECEIVER. THE DPDK RECEIVE RATE IS SHOWN IN THE TABLE AS AN APPROXIMATION OF THE ACTUAL SEND RATE.

|  | Estimated and actual send rates | | | | | | | |
|---|---|---|---|---|---|---|---|---|
|  | nativeUDP | | LibPcap | | NetMap | | DPDK | |
|  | Est | Act | Est | Act | Est | Act | Est | Act |
| Min | 2.21 | 2.19 | 5.90 | 4.38 | 9.67 | 9.81 | 9.98 | 9.88 |
| Max | 2.62 | 3.13 | 7.13 | 7.53 | 9.77 | 10.10 | 9.99 | 10.56 |
| Mean | 2.45 | 2.83 | 6.69 | 6.45 | 9.72 | 9.97 | 9.98 | 10.19 |
| Std | 0.12 | 0.31 | 0.36 | 1.12 | 0.03 | 0.08 | 0.0007 | 0.26 |
| Rel std | 5 % | 11 % | 5 % | 17 % | 0.3 % | 1 % | 0.01 % | 3 % |

The results from the four experiments are shown in Table 1. For nativeUDP it can be seen that the actual send rates are on average 2.8 Gbps, varying between 2.2 and 3.1 Gbps and having a relative standard deviation around 11%. The estimated send rates are in line with the actual send rates. This low send rate can be explained by the time taken to build IP and MAC headers in the network stack, copy packets between multiple queues in the kernel stack and at the interface level plus delays from context switching and interrupts. Luigi Rizzo et al. [11] found similar results with FreeBSD: the time for a packet going from the application to the device drivers is around 4.7 micro seconds, which corresponds to 2.5 Gbps for 1500-byte packets.

The LibPcap method generates trains at around 6.5 Gbps, with a variation between 4.4 and 7.5 Gbps and a relative standard deviation of 17%. Also for this method the estimated send rates seem to be on par with the actual send rates. This doubled send rate of LibPcap compared to nativeUDP in mainly due to manually building the headers and injecting packets at the Ethernet level.

Netmap generates trains at full speed, reporting estimated send rates at 9.7 Gbps. This result is due to pre-allocated memory buffers for packets and mapping interface directly with the user-application, thereby completely by-passing queues, in the same way as DPDK does. DPDK is also shown for reference and performs similar to Netmap with trains generated at full speed with a relative standard deviation of around 2%.

The conclusion from this set of experiments is that both Netmap and DPDK can be used as senders for 10 Gbps measurements, with very high accuracy and precision of the actual send rate and with reported estimated send rates very close to the actual rates. For measurement scenarios up to 6 Gbps the LibPcap method may be used as a sender and for link speeds up to 2.5 Gbps nativeUDP is also an alternative, if 11 - 17% relative standard deviation is acceptable precision.

*D. Comparing Receiver methods with DPDK as Sender*

The fourth set of experiments is using DPDK as a Sender and looks into the estimated receive rates for each of the four methods as the Receiver. The actual send rate from the DPDK sender is 10 Gbps, and it is assumed that these send rates are equal to the actual receive rates in the receiver, since the link is empty.

TABLE 2. ESTIMATED RECEIVE RATES IN GBPS FOR DIFFERENT METHODS AS RECEIVERS. DPDK IS SENDER IN ALL CASES AND SENDS PACKET TRAINS WITH ACTUAL SEND RATE OF 10 GBPS.

|  | Estimated receive rates | | | |
| --- | --- | --- | --- | --- |
|  | nativeUDP | LibPcap | NetMap | DPDK |
| Min | 5.09 | 4.88 | 9.63 | 9.82 |
| Max | 10.65 | 6.21 | 461.14 | 10.50 |
| Mean | 8.47 | 5.47 | 119.31 | 10.02 |
| Std | 1.70 | 0.39 | 186.96 | 0.19 |
| Rel std | 20 % | 7 % | 157 % | 2 % |

Table 2 shows the results from the experiments. It is clear that DPDK is the only method that can estimate the receive rate accurately. The nativeUDP method shows rates that are only 10-15% below the actual receive rates, which makes it interesting for scenarios up to 8 Gbps. However, it has a relative standard deviation around 20%, which probably makes it less interesting as a receiver at these high rates.

LibPcap has the lowest receiving rate of 5.5 Gbps. One possible reason is the fact the entire packet including all headers is copied to the user space and the headers are processed manually.

Netmap shows unnatural high receive rates, with a mean of 119 Gbps, which is even higher than in the first set of experiments. This dubious effect is obviously the result of that the packets are buffered before time-stamping, as illustrated in Fig. 2.

## VI. DISCUSSION

The evaluation of the four methods shows that each method has its pros and cons. It is clear that DPDK is the only method that fulfills the requirements of generating and receiving packet trains at full link speed of 10 Gbps accurately and with high precision. On the downside, all packet headers need to be provided by the application. This can be tricky to achieve, since the Ethernet header of the receiver is not commonly known to the user application at the server side. This must therefore be solved in a real measurement system, perhaps with some connection to the ARP functionality. Also, the DPDK system locks the NIC interface during the whole measurement session, which means that the interface totally disappears from the IP stack of the host. This means that no other traffic can use this interface of the server. Specifically a DPDK receiver cannot respond to ARP messages, which means that it is hard for the sender to detect the Ethernet address of the receiver. One solution to this might be to split the traffic from the interface in the DPDK receiver implementation and redirect the non-measurement traffic to a virtual interface in the operating system for further processing. Intel provides a driver for this, so it seems to be possible. Another drawback of DPDK is that CPU usage is high, due to continuous polling during the whole measurement session. Ideally, the polling should be off while no measurement trains are arriving, but the problem is then to start polling quickly enough in order to get the time stamps correct for the first packets in the train. Future research is needed for solving this.

NativeUDP suffers from passing packets between queues in IP stack and from the latency in packet-header processing. However, while showing the lowest send rates at 2.5 Gbps, it is also the simplest method to implement. It might therefore be a useful method for desired send rates up to 2.5 Gbps. What is striking is that the reported received rate of a nativeUDP receiver is over 8 Gbps for 10 Gbps send rates. Apparently the distortion of the time stamping is not the main problem for this method, but rather the ability to generate high send rates. Future work is needed to show if nativeUDP can be used as a lightweight receiver up to 8 Gbps with low CPU usage.

LibPcap gives a maximum send rate of 6 Gbps, but on the sender side the fluctuation is very high. Also, all packet headers have to be created by the application. This method is therefore not very suitable for high-speed measurements.

Netmap shows erroneous high receive rates, while performing as well as DPDK on the sender side. Compared to DPDK, Netmap locks the NIC interface only during each measurement train, but can hand it back to other applications in between the trains. However, CPU usage is high, due to polling. Perhaps this method can be used as a sender, while some other method should be used as receiver.

The DPDK method was used as a reference tool for the other methods, since it was assumed to generate and report the correct send and receive rates. In an ideal case, a hardware-based tool should be used to verify this assumption and all the methods, e.g. an IXIA or similar, but this kind of tool was not available during the experiments.

In this study we have limited the tests to an empty link of 10 Gbps, which means that the received rate can be seen as a direct estimate of the APC of the link. In a future work we plan to do similar tests with cross traffic in the network, which require a more sophisticated algorithm for APC estimation.

## VII. CONCLUSIONS

The objective of this work was to investigate send and receive capabilities of software and hardware to enable active measurements of available capacity over high-speed links. We selected four different probe-packet handling methods and evaluated in a testbed their ability to send and receive probe trains over a 10 Gbps link.

The first finding of the experiments is that the IP stack becomes the bottleneck when using native UDP sockets and when the desired send rate is higher than around 2.5 Gbps. The main reason for this is the large processing time of the packets in the stack.

The second finding is that it is possible to overcome this limitation and achieve accurate send and receive rates for measurements on 10 Gbps links using a method based on DPDK. This method bypasses the stack completely and binds the NIC directly to the application. We show that the accuracy and precision of this method is very high. However, there is also a price for this high performance in high CPU usage and blocking of the NIC for other applications.

The third finding is that some of the four methods have properties that may be valuable for measurements with lower rates than 10 Gbps or as separate senders or receivers in a measurement system. One example is nativeUDP which is simple to implement and has limited impact on the underlying system. It may therefore be a preferable choice for measurements up to 2.5 Gbps. Our tests also indicate that it might be possible to use it as a lightweight receiver with low impact on CPU or NIC binding for receive rates up to 8 Gbps. Another example is Netmap, which can send trains at full speed with the same precision as DPDK, but cannot be used as a receiver, due to erroneous receive rates.

Future work is needed to evaluate the DPDK method for APC measurements in realistic networks with cross traffic. It is also important to work on optimization of the method in order to minimize CPU consumption and impact of NIC locking.

## REFERENCES

[1] Integra: Upward Mobility - Moving IT Business to the Cloud, http://www.integratelecom.com/resources/Assets/cloud-migration-wp-integra.pdf

[2] Ekelin. S. et al.: Real-Time Measurement of End-to-End Available Bandwidth using Kalman Filtering, IEEE Network Operations and Management Symposium (NOMS), 2006.

[3] Jain, M. and Dovrolis, C.: End-to-end available bandwidth: Measurement methodology, dynamics, and relation with TCP throughput, in Proceedings of ACM SIGCOMM, Pittsburg, PA, USA, Aug. 2002.

[4] Ribeiro, V. et al.: pathChirp: efficient available bandwidth estimation for network paths, Passive and Active Measurement workshop, 2003.

[5] International Telecommunication Union (ITU-T) Recommendation Y.1540, 2011

[6] Chaudhari, S.S., Biradar, R.C.: Survey of Bandwidth Estimation Techniques in Communication Networks, in Wireless Personal Communications: Volume 83, Issue 2, 2015.

[7] Yin, Q. et al.: Can Bandwidth Estimation Tackle Noise at Ultra-High Speeds?, in 2014 IEEE 22nd International Conference on Network Protocols, p107-118, 2014.

[8] Hedayat, K. et al.: A Two-Way Active Measurement Protocol (TWAMP). IETF RFC 5357, October 2008.

[9] Baillargeon, S. et al.: Ericsson Two-Way Active Measurement Protocol (TWAMP) Value-Added Octets. IETF RFC 6802, November 2012.

[10] Garcia, L. M.: Programming with LibPcap – Sniffing the network from our own application, Hakin9 – Computer Security Magazine 2008.

[11] Rizzo, L.: netmap: a novel framework for fast packet I/O, Proceedings of the USENIX Annual Technical Conference, 2012.

[12] Intel DPDK: Data Plane Development Kit, http://dpdk.org/

[13] Emmerich, P. et al.: MoonGen: A Scriptable High-Speed Packet Generator, 2015, http://arxiv.org/ftp/arxiv/papers/1410/1410.3322.pdf